\documentclass[12pt]{article}
\usepackage{amsfonts,amssymb,amsmath,amscd,theorem,oldgerm}
\usepackage[mathscr]{eucal}

\newtheorem{lemma}{Lemma}[section]
\newtheorem{proposition}[lemma]{Proposition}
\newtheorem{remark}[lemma]{Remark}

\newtheorem{theorem}[lemma]{Theorem}

\newtheorem{definition}[lemma]{Definition}
\newtheorem{corollary}[lemma]{Corollary}

\def\eps{\varepsilon}

\def\ome {\omega}

\def\lam{\lambda}
\def\Lam{\Lambda}
\def\G{\mathrm{G}}
\def\A{{\cal A} _{\varepsilon,\hbar}}

\def\Wh{W_\h}
\def\rhoh{\rho _{_\h}}

\def\pih{\pi _{_\h}}
\def\lto{\longrightarrow}
\def\lang{\langle}
\def\rang{\rangle}
\def\h{\hbar}
\def\F{C^{\infty}({\bf T})}
\def\rev{\quad}
\def\R{{\mathbb R}}
\def\Q{{\mathbb Q}}
\def\N{{\mathbb N}}
\def\Z{{\mathbb Z}}
\def\C{{\mathbb C}}
\def\T{{\bf T}}

\def\r{\;}

\def\t{\text}

\def\cent{Z}

\def\iso{\backsimeq}
\def\Irr{\mathrm{Irr}}
\def\Sp{\mathrm{Sp}}
\def\f{\mathrm{f}}
\def\2n{\mathrm{2n}}
\def\n{{\bf n}}
\def\m{{\bf m}}

\def\x{{\bf x}}
\def\L{\mathrm{L}}
\def\TC{\mathbb{T}(\C)}

\newcommand{\eProof}{\hfill \square}

\begin{document}

\title{\bf The Multidimensional Berry-Hannay Model}

\author{S. GUREVICH $\r$  and  R. HADANI  \\
  School of Mathematical Sciences \\
 Tel-Aviv University}

\maketitle

\bigskip

\begin{abstract}
The aim of this paper is to construct the Berry-Hannay model of
quantum mechanics on a 2n-dimensional symplectic torus. We
construct a simultaneous quantization of the algebra $\cal A$ of
functions on the torus and the linear symplectic group $\G =
\Sp(\mathrm{2n},\Z)$. In the construction we use the quantum torus
$\A$, which is a deformation of $\cal A$, together with a
$\G$-action on it. We obtain the quantization via the action of
$\G$ on the set of equivalence classes of irreducible
representations of $\A$. For $\h \in \Q$ this action has a unique
fixed point. This gives a canonical projective equivariant
quantization. There exists a Hilbert space on which both $\G$ and
$\A$ act in a compatible way.
\end{abstract}

\bigskip

\setcounter{section}{-1}

\section{Introduction}

\subsection{Motivation} In the paper ``{\it Quantization of linear maps on the torus - Fresnel
diffraction by a periodic grating}'', published in 1980 (see
\cite{BH}), the physicists M.V. Berry and J. Hannay explore a model
for quantum mechanics on the 2-dimensional torus. One of the
motivations was to study the phenomenon of quantum chaos in this
model (see \cite{R} for a survey). \\

Berry and Hannay suggested to quantize simultaneously the functions on the torus
and the linear symplectic group $\mathrm{Sp}(2,\Z) = \mathrm{SL}(2,\Z)$.

In this paper we want to extend our construction \cite{GH} of the
2-dimensional Berry-Hannay model to the higher dimensional tori.
The central question is whether there exists a vector space on
which a deformation of the algebra of functions and the linear
symplectic group $\Sp(\2n,\Z)$, both act in a compatible way.
Previously it was shown by  Bouzouina  and De Bievre (see
\cite{BDB}) that one can quantize simultaneously the functions on the torus
and one ergodic element $A \in \mathrm{Sp}(\2n,\Z)$ in case when the Planck
constant is of the form $\h = \frac{1}{N},\; N\in \N $.

\subsection{Definitions}
\subsubsection{Equivariant quantization of the torus}\label{DefQ}
Fix a 2n-dimensional vector space $V$ over $\R$ and a rank 2n
lattice $\Lam \subset V$. Let $\T = V/\Lam $ be the 2n-dimensional
torus determined by $\Lam$. We fix a skew-symmetric bilinear form
$\ome$ on $V$, which we consider as a differential form on $\T$;
we assume that vol$(\T)=1$.

Denote by $\G$ the group $\mathrm{Sp}(\T)$ of all the
automorphisms of $V$ which takes $\Lam$ into itself and preserve
the form $\ome$.

Let ${\cal A} = \F$ be the algebra of smooth complex valued functions on $\T$,
and let $\Lam ^* := \{ \xi \in V^* | \; \xi(\Lam) \subset \Z \} $
be the dual lattice of $\Lam$. We use the lattice $\Lam^*$ in
order to have a canonical basis for ${\cal A}$. Let $<,>$ be the pairing between $V$ and $V^*$. The map $\xi
\mapsto s(\xi)$ where $s(\xi)(x) := e^{2\pi i <x,
\xi>},\r x \in \T,\r \xi\in \Lam^*$ defines a canonical
isomorphism between $\Lam^*$ and the group $\T^* := \mathrm{Hom} (\T,\C^\times)$
of characters of $\T$. By Fourier theory the last group form a
basis to ${\cal A}$.
\\\\

We will construct  a particular type of quantization procedure for
the functions (see also \cite{GH}). Moreover this quantization
will be equivariant with respect to the action of the group of
``classical symmetries'' $\G$:

\begin{definition} A $\r${\bf Weyl quantization} of ${\cal A}$ is a family of
$\C$-linear morphisms $\pi _{_\h}:{\cal A} \lto
\mathrm{End}(W_\h), \r \h \in \R$ s.t the following property
hold:
\begin{equation}
\pi _{_\h} (s(\xi))\pi _{_\h} (s(\eta)) = e^{2 \pi i \h
\ome(\eta,\xi)} \pi _{_\h}(s(\eta)) \pi _{_\h}(s(\xi))
\end{equation}
for all $\xi,\eta \in \Lam ^*$ and $\h \in \R$. Here the form $\ome$ denote the form on $V^*$ induced by the symplectic form
on $V$.
\end{definition}

\begin{definition} An {\bf{equivariant quantization}} of $\T$ is a quantization of $\cal A$ with additional maps
$\rhoh:\G \lto  \mathrm{GL} (W _\h) \rev $ s.t. the following
equivariant property $($called ''Egorov
identity''$)$ holds:
\begin{equation}\label{Egorov}
{\rho _{_\h}  (B)}^{-1} \pi _{_\h}(f) \rho _{_\h} (B) = \pi
_{_\h}(f\circ B)
\end{equation}
for all $\h \in \R , \r f \in {\cal A} \r$ and $\r B \in \G$. If
$(\rho _{_\h},W _\h) $ is a projective representation of the
group $\G$ then we call the quantization {\bf projective}.
\end{definition}

\subsection{Results}
In this paper we give an affirmative answer to the existence of
the quantization procedure and give explicit formulas. We show a
construction (Theorem \ref{GH}, Corollary \ref{prq}) of the
quantization procedure for rational Planck constants. This is the
first construction of such equivariant quantization for higher
dimensional tori.
\\

The idea of the construction is as follows: We use a "deformation"
of the algebra $\cal A$ of functions on $\T$. We define (see
\ref{qs}) two algebras $\A, \r \eps = 0,1$. The algebra ${\cal A}
_{0,\hbar}$ is the usual Rieffel's quantum torus (see \cite{Ri})
and ${\cal A} _{1,\hbar}$ is some twisted version of it. If $\h =
\frac{M}{N} \in \Q$, then we will see that all irreducible
representations of $\A$ have dimension $N^n$. We denote by
$\Irr(\A)$ the set of equivalence classes of irreducible algebraic
representations of the quantized algebra. We will see that
$\Irr(\A)$ is a set "equivalent" to a torus.

The group $\G$ naturally acts on a quantized algebra $\A$ and
hence on the set $\Irr(\A)$.

Let $\h = \frac{M}{N}$ with $\mathrm{gcd}(M,N) =1$. Set
$\varepsilon = MN$ (mod 2). Then:

\begin{theorem}[Canonical equivariant representation]\label{GH}  There exist a {\bf unique} $($up to
    isomorphism$)$ irreducible representation
    $(\pi,W)$ of $\A$ for which its equivalence class is fixed by
$\G \r($i.e $\r \pi  \iso  {}^B\pi$ for all $B \in \G)$.
\end{theorem}

We will give formulas for the canonical representation in section
\ref{F}.
\\

Since the canonical representation $(\pi,W)$ is
irreducible, by Schur's lemma we get the canonical projective representation of $\G$ compatible with $\pi$:

\begin{corollary}[Canonical projective representation]\label{prq}
For every $B \in \G$ there exist an operator $\rho(B)$ on $W$
s.t:
\begin{equation}
{\rho (B)}^{-1} \pi (f) \rho (B) = \pi (f\circ B) \rev \rev
\text{for all} \r f \in {\cal A}
\end{equation}
Moreover the correspondence $B \mapsto \rho(B)$ gives a projective
representation of $\G$.
\end{corollary}

We will give formulas for the projective representation in  section \ref{F}.
\\

\begin{remark} The family $(\rhoh,\pih,\Wh), \rev \h \in \Q$ presented in Corollary \ref{prq}
gives a canonical equivariant quantization of the torus.
We can endow $($see $\ref{us})$ the space $\Wh$ with a canonical unitary
structure s.t $\pih$ is a $*$-representation and $\rhoh$ unitary.
This answer the question whether this quantization is also unitarizable and hence fits to the idea that quantum
mechanics should be realized on a Hilbert space.
\end{remark}

\subsection*{Acknowledgments}
We thank warmly our Ph.D. adviser J. Bernstein for his interest
and guidance in this project. We thank very much P. Kurlberg and
Z. Rudnick who discussed with us their papers and explained their
results. Special thanks to M. Tiecher. This paper was written during our
visit to work with Prof. P. Kurlberg at Chalmers university,
Gothenburg, Sweden at Sep. 03. We thank Prof. Kurlberg for his
invitation and the mathematics department at Chalmers university
for the hospitality.

The work of S.G was partially supported by the European Commission
under the Research Training Network (Mathematical Aspects of
Quantum Chaos).

\section{Construction}\label{Con}
Set ${\cal A} = \F$ the algebra of smooth complex valued function on the torus. Let $\Lam ^* := \{ \xi \in V^* | \; \xi(\Lam)
\subset \Z \} $. Let $<,>$ be the pairing between $V$
and $V^*$. The map $\xi \mapsto s(\xi)$ where $s(\xi)(x) :=
e^{2\pi i <x,\xi>},\r x\in \T,\r \xi\in \Lam^*$ define a
canonical isomorphism between $\Lam^*$ and the group $\T^*  :=
\mathrm{Hom} (\T,\C^\times)$ of characters of $\T$.

\subsection{The quantum tori}\label{qs}
Fix $\h \in \R.$ Define two algebras (see also \cite{Ri} and
\cite{GH})) $\A,\r \eps = 0,1$ as follows. The algebra $\A$ is
defined over $\C$ by generators $\{s(\xi),\r \xi \in \Lam ^* \}$,
and relations:
\begin{equation}\label{qt}
s(\xi+\eta) = (-1)^{\eps\ome(\xi,\eta)}e^{\pi i \h
\ome(\xi,\eta)} s(\xi)s(\eta)
\end{equation}
for all $\xi,\eta \in \Lam^*$.

\subsection{Weyl quantization}
To get a Weyl quantization of $\cal A$ we use a specific one-parameter family of representations (see subsection
\ref{ceq} below) of the quantum tori.
This define an operator $\pi(s(\xi))$ for every $\xi \in \Lam^*$.
We extend the construction to every function $f \in {\cal A}$
using Fourier theory. Suppose
\begin{equation}
f = \sum\limits_{\xi \in \Lam^*} a_{_\xi}
s(\xi)
\end{equation}
is its Fourier expansion. Then we define its {\it Weyl quantization} by
\begin{equation}
\pi(f) = \sum\limits_{\xi \in \Lam^*} a_{_\xi} \pi(s(\xi))
\end{equation}
The convergence of the last series is due to the rapid decay of the fourier
coefficients of $f$.

\subsection{Equivariant quantization}
We describe a strategy how to get an equivariant quantization of
$\T$. The group $\G = \Sp(\T)$ acts on $\Lam$ preserving
$\ome$. Hence $\G$ acts on $\A$. The formula of this action is
${}^B s(\xi):= s(B\xi)$. Suppose $(\pi,W)$ is a representation of
$\A$. For an element $B \in \G$, define ${}^B\pi(s(\xi)) :=
\pi({}^{B^{-1}}s(\xi))$. This formula defines an action of $\G$ on
the set $\Irr(\A)$ of equivalence classes of irreducible algebraic
representations of $\A$.

\begin{lemma}\label{dim} All irreducible representations of $\A$
are $\mathrm{N}^n$-dimensional.
\end{lemma}

Suppose $(\pi,W)$  is a representation for which its equivalence
class is fixed by the action of $\G$. This means that for any
$B\in \G$ we have $\pi \iso {}^B\pi$, so by definition there exist
an operator $\rho(B)$ on $W$ s.t:
\begin{equation}
{\rho(B)}^{-1} \pi(s(\xi)) \rho(B) = \pi(s(B\xi)), \rev \t{for
all} \r \xi\in \Lam ^*
\end{equation}
This imply the Egorov identity (\ref{Egorov}) for any function.
Suppose in addition that $(\pi,W)$ is an irreducible representation. Then by
Schur's lemma for every $B\in \G$ the operator $\rho(B)$ is
uniquely defined up to a scalar. This implies that $(\rho,W)$ is
a projective representation of $\G$.

\subsection{The canonical equivariant quantization}\label{ceq}
In what follows we consider only the case  $\h \in \Q$. We write
$\h$ in the form $\h = \frac{M}{N}$ with $\mathrm{gcd}(M,N) =1$.
Set $\varepsilon = MN$ (mod 2).

\begin{proposition}\label{ufp}
There exist a unique $\pi \in \Irr(\A)$ which is a fixed point for
the action of $\G$.
\end{proposition}

\subsection{Unitary structure}\label{us}
Note that $\A$ becomes a $\r *-$ algebra by the formula $s(\xi)^*
:= s(-\xi)$. Let $(\pi,W)$ be the canonical representation of
$\A$.

\begin{proposition}\label{us}
There exist a canonical $\,($unique up to scalar$)$ unitary
structure on $W$  for which $\pi$ is a $*-$representation.
\end{proposition}

\section{Formulas}\label{F}
We give formulas for the canonical projective equivariant quantization $(\pi,\rho,W)$. This type of realization is called
the {\it Schrodinger model}.
\\

Choose a polarization $\Lam^* = \L \oplus \L'$ with $\ome|\L =
\ome|\L' = 0$. The form $\ome$ defines a non-degenerate pairing
between $\L$ and $\L'$ given by $<\xi,\xi'> =
\ome(\xi,\xi')$. Set $X = \L/N\L$ and $W = {\cal F}(X)$ the
space of functions on $X$. Denote by $\psi$ the additive character
of $\Z/N\Z$ given by $\psi(t) = e^{2 \pi i \h t}$.

\subsection{Coordinates free formulas}

\subsubsection{Formula for $\pi$}
The representation $\pi$ is given by:
\begin{equation}
\left[\pi(\xi)\f\right](x) = \f(x+\xi)
\end{equation}
\begin{equation}
\left[\pi(\xi')\f\right](x) = \psi(<x,\xi'>)\f(x)
\end{equation}
for every $\xi \in \L, \; \xi' \in \L',\; x \in X$ and $\f \in {\cal F}(X)$.
\\

The formula for general element is given by:

\begin{equation}
\left[\pi(\xi + \xi')\f\right](x) = \alpha(\xi,\xi') \psi(<x,\xi'>)\f(x+\xi)
\end{equation}
\\

where $\alpha (\xi,\xi') := (-1)^{\varepsilon<\xi,\xi'>}e^{\pi i \h <\xi,\xi'>}$.

\subsubsection{Formula for $\rho$}\label{frho}
The projective representation $\rho$ is described by the following formulas:

\begin{equation}
\left[\rho\left(
\begin{array}{cc}
 A & 0 \\
 0 & ^tA^{-1}
\end{array} \right)\f\right](x) = \f(A^{-1}x)
\end{equation}
for every $A \in \mathrm{GL}(\L)$ - the group of invertible operators which preserve the lattice $\L$.
Here $^tA$ denote the operator on $\L'$ dual to $A$.

\begin{equation}
\left[\rho\left(
\begin{array}{cc}
 I & 0 \\
 B & I
\end{array} \right)\f\right](x) = (-1)^{M<x,Bx>}e^{\pi i \h
<x,Bx>}\f(x)
\end{equation}
for every $B:\L \lto \L'$ symmetric bilinear form.
\\

And for every non-degenerate symmetric bilinear form $B:\L \lto \L'$,
\begin{equation}
\left[\rho\left(
\begin{array}{cc}
 0 & -B^{-1} \\
 B & 0
\end{array} \right)\f\right](x) = \widehat{\f}(x)
\end{equation}
where $\widehat{\f}$ denote the Fourier transform
$\widehat{\f}(x) := \frac{1}{\sqrt{N^n}} \sum\limits_{y \in
X} \f(y) \psi (<x,By>)$.

\subsection{Formulas with coordinates}
Choose bases $(e_1,\ldots,e_n)$ and $(e_1',\ldots,e_n')$ for $\L$
and $\L'$ respectively s.t $\ome(e_i,e_j') = \delta_{ij}$. This
allows us to identify $\Lam ^*$ with $\Z^n \oplus \Z^n$, the set $X$
with $\Z^n/N\Z^n$ and the
group $\G = \mathrm{Sp}(\T)$ with $\mathrm{Sp}(\2n,\Z)$.

\subsubsection{Formula for $\pi$} The representation $\pi$ is given by:
\begin{equation}
\left[\pi(\m,\n)\f\right](\x) = \alpha(\m,\n) \psi(<\x,\n>)\f(\x+\m)
\end{equation}
\\
for every $\m \in \L, \; \n \in \L',\; \x \in X$ and $\f \in {\cal F}(X)$,
where $\alpha (\m,\n) := (-1)^{\varepsilon<\m,\n>}e^{\pi i \h <\m,\n>}$.

\subsubsection{formula for $\rho$} The projective representation $\rho$
is described by the following formulas:

\begin{equation}
\left[\rho\left(
\begin{array}{cc}
 A & 0 \\
 0 & {}^tA^{-1}
\end{array} \right)\f\right](\x) = \f(A^{-1}\x)
\end{equation}
for every $A \in \mathrm{GL}(n,\Z)$.
\begin{equation}
\left[\rho\left(
\begin{array}{cc}
 I & 0 \\
 B & I
\end{array} \right)\f\right](\x) = (-1)^{M<\x,B\x>}e^{\pi i \h
<\x,B\x>}\f(\x)
\end{equation}
for every symmetric matrix $B \in \mathrm{Mat(n,\Z)}$.
\\

And for every invertible symmetric matrix $B \in \mathrm{Mat(n,\Z)}$:
\begin{equation}
\left[\rho\left(
\begin{array}{cc}
 0 & -B^{-1} \\
 B & 0
\end{array} \right)\f\right](\x) = \widehat{\f}(\x)
\end{equation}
where $\widehat{\f}$ denote the Fourier transform as defined in \ref{frho}.

\section{Proofs}\label{P}

\subsection{Proof of Proposition \ref{dim}} Suppose $(\pi,W)$ is an
irreducible representation of $\A$.

{\it Step 1.} First we show that $W$ is finite dimensional. The algebra $\A$
is a finite module over $\cent (\A) = \{s(N\xi),\r \xi \in \Lam ^*
\}$ which is contained in the center of $\A$. Because $W$ has at
most countable dimension (as a quotient space of $\A$) and $\C$ is
uncountable then by Kaplansky's trick (See \cite{MR}) $\cent (\A)$
acts on $W$ by scalars. Hence dim $W < \infty $.

{\it Step 2.} We show that $W$ is $\mathrm{N}^n-$dimensional.

Choose a basis $(e_1,\ldots,e_n,e_1',\ldots,e_n')$ of $\Lam^*$ s.t
$\ome(e_i,e_j) = \ome(e_i',e_j')= 0$ and $\ome(e_i,e _j') =
\delta_{ij}$ the Kronecker's delta. Denote by $E$ the commutative
subalgebra of $\A$ generated by $\{s(e_i)\}_1^n$. Suppose $\lam
\in E^*$ is an eigencharacter of $E$ and denote by $W _\lam =
W_{(\lam_1,\ldots,\lam_n)}$ the corresponding eigenspace, $\lam_i
:= \lam(e_i)$. We have the following commutation relation $\pi
(e_i) \pi (e_j') = \gamma^{\delta_{ij}} \pi (e_j') \pi (e_i)$
where $\gamma := e^{-2 \pi i \frac{M}{N}}$. Hence $\pi (e_j'): W
_{(\lam_1,\ldots,\gamma ^k\lam_j,\ldots,\lam_n)} \lto W
_{(\lam_1,\ldots,\gamma ^{k+1}\lam_j,\ldots,\lam_n)}$. Since
$\mathrm{gcd}(M,N) = 1$ the eigencharacters
$(\gamma^{k_1}\lam_1,\ldots,\ldots,\gamma^{k_n}\lam_n), \r 0 \leq
k_j \leq N-1$ are all different. Let $0 \neq w \in W _\lam$.
Recall that $\pi (e_j') ^N = \pi(N e_j')$ is a scalar operator. The
space $\mathrm{span}\{\pi(e_j')^{k}w\}$ is $N^n$-dimensional $\A
-$invariant subspace hence equal $\r W$. $\eProof$

\subsection{Proof of Proposition \ref{ufp}} Suppose $(\pi,W)$  is an
irreducible representation of $\A$.  By Schur's lemma for every
$\xi\in \Lam^*$, $\pi(N\xi)$ is a scalar operator $\pi (N\xi) =
\chi_{_\pi}(\xi) \cdot \mathrm{I}$. We have $\pi(0) = \mathrm{I}$, hence $\chi_{_\pi}(\xi) \neq 0$ for all $\xi \in \Lam^*$. Thus to any irreducible
representation we have attached a scalar function $\chi
_{_\pi}:\Lam ^* \lto \C^\times$. It is easy to see that
$\chi_{_\pi}(\xi+\eta) = \chi_{_\pi}(\xi)\chi_{_\pi}(\eta)$. Let
$\TC := \mathrm{Hom} (\Lam ^*,\C^\times)$ be the group of complex
characters of $\Lam ^*$. We have defined a map $\Irr(\A) \lto
\TC$ given by $\pi \mapsto \chi _{_\pi}$. This map is obviously
compatible with the action of $\G$, where the group $\G$ acts on
characters by $\r {}^B\chi(\xi):=\chi(B^{-1}\xi)$.

\begin{lemma}\label{realiz}
The map $\pi  \mapsto \chi _{_\pi}$ gives a $\G$-equivariant
bijection:
\[
\begin{CD}
\mathrm{Irr}(\A)    @>  >>  \TC
\end{CD}
\]
\end{lemma}

From Lemma \ref{realiz} we easily deduce that there exists a
unique equivalence class $\pi \in \Irr(\A)$ which is fixed by the
action of $\G$. This is the one that corresponds to the trivial
character ${\bf 1} \in \TC$ which is the unique fixed point for
the action of $\G$ on $\TC$.

{\bf Proof of Lemma \ref{realiz}.} {\it Step 1.} The map $\pi
\mapsto \chi _{_\pi}$ is onto. We define an action of $\TC$ on
$\Irr(\A)$ and on itself by $\pi \mapsto \chi\pi$ and $\psi\mapsto
\chi^N \psi$, where $\chi \in \TC$, $\pi \in \Irr(\A)$ and
$\psi \in \TC$. The map $\pi \mapsto \chi _{_\pi}$ is clearly a
$\TC$-equivariant map with respect to these actions. The claim
follows since the above action of $\TC$ on itself is
transitive.

{\it Step 2.} The map $\pi \mapsto \chi _{_\pi}$ is one to one.
Suppose $(\pi,W)$ is an irreducible representation of $\A$. It is
easy to deduce from the proof of Lemma \ref{dim} that for $\xi
\notin N\Lam ^*, \r \mathrm{tr}(\pi(\xi)) = 0$. But we know from
character theory that an isomorphism class of a finite dimensional
irreducible representation of an algebra is recovered from its
character. $\eProof$

\subsection{Proof of Proposition \ref{us}}

{\it Existence.} We use the realization of the representation $\pi$ given in section \ref{F}. It is easy to see that
the representation
$(\pi,{\cal F}(X))$ is a $*-$representation with respect to the standard scalar product
on ${\cal F}(X)$ defined by:
\begin{equation}
\lang \f,g \rang = \frac{1}{\mathrm{N}^n} \sum\limits_{x \in X} \f(x)\overline{g(x)}
\end{equation}

{\it Uniqueness.} From Schur's lemma follows that the Hilbert structure
endowed on the vector space of the representation is unique up to
a scalar.

$\eProof$

\begin{remark} It can be shown that all the
irreducible $*$-representations of $\A$ on Hilbert spaces are the
ones for which $\chi_{_\pi}$ takes values in the circle $S^1$.
\end{remark}

\bigskip \noindent
Gurevich Shamgar\\
School of Mathematical Sciences \\
Tel Aviv University \\
Tel Aviv 69978, Israel \\
{\it e-mail: gorevitc@post.tau.ac.il}

\bigskip \noindent
Hadani Ronny\\
School of Mathematical Sciences\\
Tel Aviv University \\
Tel Aviv 69978, Israel \\
{\it e-mail: hadani@post.tau.ac.il}

\begin{thebibliography}{R}


\bibitem[BH]{BH}Hannay J.H and Berry M.V, Quantization of linear maps on the torus -
Fresnel diffraction by a periodic grating. {\it Physica D 1} (1980), 267-291.

\bibitem[BDB]{BDB} Bouzouina A., De Bievre S., Equipartition of the eigenfunctions of quantized ergodic maps on the torus.
{\it Commun.Math.Phys. 178} (1996), 83-105.

\bibitem[GH]{GH} Gurevich S.  and  Hadani R., On Berry-Hannay Equivariant Quantization
of the Torus. arXiv:math-ph/0312039.

\bibitem[MR]{MR} McConnell J.C. and Robson J.C., Non-commutative Noetherian Rings. {\it Graduate studies in
mathematics, 30. Amer. Math. Soc. Providence, RI} (2001), 343-344.

\bibitem[R]{R} Z. Rudnick, On quantum unique ergodicity for linear maps of the torus. To appear
in {\it Proceedings of the third European Congress of Mathematicians July 2000, Barcelona}.

\bibitem[Ri]{Ri} Rieffel, M. A. Non-commutative tori---a case study of non-commutative differentiable manifolds,
{\it Contemporary Math. 105} (1990), 191-211.
\end{thebibliography}
\end{document}